\documentclass[aps,nofootinbib,showpacs,showkeys,preprint
,tightenlines ] {revtex4}

\usepackage{epsf,epsfig,subfigure,graphicx,amsmath,amssymb}
\usepackage{color}

\newcommand{\dis}[1]{\begin{equation}\begin{split}#1\end{split}\end{equation}}
\newcommand{\be}{\begin{equation}}
\newcommand{\ee}{\end{equation}}

\newcommand{\eq}[1]{Eq.~(\ref{#1})}

\newcommand{\bfrac}[2]{{\left(\frac{#1}{#2} \right)  }}
\newcommand{\Mp}{M_P}

\newcommand{\gev}{\,\textrm{GeV}}

\newcommand{\calP}{{\cal P}}

\def\bea{\begin{eqnarray}}
\def\eea{\end{eqnarray}}

\begin{document}

 \begin{flushright}
{\tt  
}
\end{flushright}

\title{\Large\bf Primordial gravitational wave of BICEP2  
\\from dynamical double hybrid inflation}

\author{Ki-Young Choi$^{(a)}$\footnote{email: kiyoungchoi@kasi.re.kr}
and Bumseok Kyae$^{(b)}$\footnote{email: bkyae@pusan.ac.kr} }
\affiliation{$^{(a)}$ Korea Astronomy and Space Science Institute, Daejon 305-348, Republic of Korea\\
$^{(b)}$ Department of Physics, Pusan National University, Busan
609-735, Republic of Korea}


\begin{abstract}
BICEP2 has observed a primordial gravitational wave corresponding to the tensor-to-scalar ratio of 0.16. 
It seems to require a super-Planckian inflationary model. 
In this paper, we propose a double hybrid inflation model, where the inflaton potential dynamically changes with the evolution of the inflaton fields. 
During the first phase of inflation over 7 e-folds, 
the power spectrum can be almost constant by a large linear term in the hybrid potential, 
which is responsible also for the large tensor-to-scalar ratio. 
In the second phase of 50 e-folds, the dominant potential becomes dynamically changed to the logarithmic form as in the ordinary supersymmetric hybrid inflation, which is performed by the second inflaton field. 
In this model, the sub-Planckian field values ($\sim 0.9~\Mp$) can still yield the correct cosmic observations with the sufficient e-folds.
\end{abstract}

\pacs{98.80.Cq, 12.60.Jv, 04.65.+e}
  \keywords{primordial gravitational wave, tensor-to-scalar ratio, spectral index, hybrid inflation}
 \maketitle


\section{Introduction}

The generation of the large scale structures and the anisotropy in the temperature of the cosmic microwave background  (CMB) suggests that there were already small inhomogeneities in the early Universe, a few Hubble times before the observable scale enters the horizon~\cite{review}.  
The time-independent  curvature perturbation $\zeta$ sets the initial conditions for such inhomogeneity and the subsequent evolution of the scalar perturbation.  
After the first observation by Penzias and Wilson (1965) fifty years ago, the precise observations of the CMB~\cite{COBE,Komatsu:2010fb,Ade:2013zuv} found that the primordial power spectrum is  Gaussian with the size of ${\cal P}_\zeta \approx 2.43\times 10^{-9}$ and is almost scale-independent with the spectral index $n_\zeta \approx 0.96$. 

The inflation models not only explain the problems of the standard big bang cosmology such as the flatness, horizon and monopole problems but also predicts the cosmological  perturbations in the matter density and spatial curvature, which explain well the primordial power spectrum~\cite{InflationFluctuation}. Those have arisen naturally from the vacuum fluctuations of light scalar field(s) during inflation, and been promoted to classical one around the time of the horizon exit. As well as the scalar perturbation, the tensor perturbation is also generated during inflation  and shows  particular features in the B-mode of the CMB polarization data. This B-mode polarization from the primordial tensor spectrum has been searched for a long time as a signature of the primordial inflation.

Recently, BICEP2~\cite{Ade:2014xna} has announced that they have measured the B-mode from the  primordial gravitational wave as well as  that from the gravitational lensing effect. The observation prefers to the non-zero tensor spectrum with the tensor-to-scalar ratio,
\dis{
r=0.2^{+0.07}_{-0.05}.
}
After foreground subtraction with the best dust model, however, the tensor-to-scalar ratio shifts down to~\cite{Ade:2014xna}
 \dis{
r=0.16^{+0.06}_{-0.05}.
}
Such a large gravitational wave has profound implications for inflation models.
The tensor power spectrum comes from the expansion of the Universe during inflation 
\dis{
{\cal P}_T = \frac{8H_*^2}{4\pi^2},
}
where $H_*$ is the expansion rate at the horizon exit, 
and thus the tensor-to-scalar ratio is given by~\cite{Liddle:1992wi} 
\dis{
r =\frac{\calP_T}{\calP_{\zeta}}=\frac{8\calP_*}{\Mp^2\calP_{\zeta}} .  
}
Here $M_P$ denotes the Planck mass ($\approx 2.4\times 10^{18} \gev$).  
Combining with the observed power spectrum~\cite{Ade:2013zuv}
\dis{
{\cal P}_\zeta = (2.198\pm0.056)  \times 10^{-9},
}
the observed large tensor spectrum corresponds to the Hubble expansion parameter
\dis{
H_*\approx 1.0\times 10^{14}\gev,
}
or to the potential energy during slow-roll  inflation
\dis{
V^{1/4}\approx 2.08\times 10^{16}\gev.
} 

However, the slow-roll condition during inflation gives the relation between the field variation and the tensor spectrum known as Lyth bound~\cite{Lyth:1996im},
\dis{
\frac{\Delta \phi}{\Mp} \gtrsim \mathcal{O} (1) \times \bfrac{r}{0.1}^{1/2} .
}
Thus, a large tensor is possible only for a large field variation, which is usually larger than the Planck scale.
More accurate bounds were studied in~\cite{lyth2,Choudhury:2013iaa} for the single field inflation.
The problem of sub-Planckian inflation with $\epsilon \approx 0.01$ is that the  e-folding number is connected to the field variation as
\dis{
\Delta N \approx \frac{1}{\Mp} \int \frac{d\phi}{\sqrt{2\epsilon}} \approx 7 ~\bfrac{\Delta \phi}{\Mp} \sqrt{\frac{0.01}{\epsilon}},
}
and so only  $\Delta N \sim 7$ is maximally obtained for $\Delta \phi\sim M_P$. 
In order to achieve a large enough e-foldings, hence, $\epsilon$ should somehow be made decreasing 
after about 7 e-folds. 
To be consistent with the observation of CMB, moreover, 
the power spectrum should be maintained as almost a constant 
even under such a large field variation for the first 7 e-folds \cite{Bringmann:2011ut}
corresponding to the observable scales by CMB, $10 ~\rm{Mpc} \lesssim k^{-1} \lesssim 10^4 ~\rm{Mpc}$~\cite{Ade:2013zuv}.  
There are some ways suggested to  accommodate the large tensor-to-scalar ratio in the sub-Planckian 
inflation models by non-monotonic evolution~\cite{Hotchkiss:2008sa,BenDayan:2009kv,Shafi:2010jr,Rehman:2010wm,Okada:2011en,Civiletti:2011qg,Hotchkiss:2011gz,Choudhury:2013jya,Antusch:2014cpa} in the single field models or in the assisted inflation~\cite{Liddle:1998jc,Kim:2006ys}. 

In the inflation with multiple scalar fields~\cite{Polarski:1992dq,Bassett:2005xm,Wands:2007bd}, however, the simple relation in the single field inflation is modified due to the  quite different inflationary dynamics. The curvature perturbation continues the evolution until the non-adiabatic perturbation is converted to the adiabatic one~\cite{Komatsu:2008hk,Choi:2008et}. 
Even the condition ending the inflation can generate  the power spectrum~\cite{Lyth:2005qk,Lyth:2006nx,Sasaki:2008uc,Naruko:2008sq,Huang:2009xa,Clesse,Byrnes:2008zy,Huang:2009vk,Yokoyama:2008xw,Emami:2011yi,Choi:2012hea}  and, therefore, changes the tensor-to-scalar ratio. However, the B-mode observation requires that the inflaton perturbation must account for much  more than 10\% of the primordial curvature perturbation for the slow-roll hypothesis~\cite{Lyth:2014yya}.


``Hybrid inflation'' \cite{Linde:1993cn} was suggested  with two scalar fields, where one is the inflaton  and the other, called the waterfall field, is to terminate inflation when it becomes tachyonic. The advantage of it is that the inflaton's field value  
is small compared to the Planck scale, and thus it is legitimate to use it as a low energy effective theory.
In the supersymmetric (SUSY) version of the hybrid inflation \cite{FtermInf2,FtermInf}, the potential can be made  flat enough, avoiding the eta-problem: fortunately the Hubble induced mass term is accidentally canceled out with the minimal K$\ddot{\rm a}$hler potential and the Polonyi type superpotential during inflation. The specific form of the superpotential can be guaranteed by the introduced U(1)$_R$ symmetry.  

By the logarithmic quantum correction to the scalar potential, the inflaton can be drawn to the true minimum, leading to reheating of the universe by the waterfall fields. Moreover, thanks to such a logarithmic correction, the vacuum expectation values (VEVs) of the waterfall fields can be determined with the CMB anisotropy \cite{FtermInf}. The VEVs turn out to be tantalizingly close to the scale of the grand unified theory (GUT). Accordingly, the waterfall fields can be regarded as GUT breaking Higgs fields in this class of models \cite{3221,422,FlippedSU(5),SO(10)}.
This inflationary model predicts a red-tilted power spectrum \cite{FtermInf} around 
\begin{equation}
n_\zeta \approx 1+2\eta\approx 1-\frac{1}{N}\approx 0.98
\end{equation}
for $N=50-60$ e-folds. It is too large compared to the present bound on the spectral index. 
At the same time, the tensor spectrum is accordingly too small to detect.
In the SUSY hybrid inflation models with a single inflaton field, it was found that the tensor-to-scalar ratio is $r\lesssim 0.03$~\cite{Shafi:2010jr,Rehman:2010wm,Okada:2011en,Civiletti:2011qg}.
 
In this paper, we study a dynamical two field hybrid inflation model \cite{twofieldhybrid}. The dominant potential changes dynamically due to the evolution of another hybrid inflaton field. In the first phase of inflation  for around $7$ e-foldings, two inflaton fields are active and generate the power spectrum. 
When the first waterfall fields are effective, one inflaton falls down to the minimum and the second phase of hybrid inflation starts. 
Since the vacuum energy and $\epsilon$ are almost constant during the first phase of inflation, 
we can obtain an almost constant power spectrum in this model. 
In the second phase of inflation, the potential has the usual shape of the logarithmic one and gives a sufficient e-folding number until the second waterfall fields are effective and the whole inflation ends. 
Since $\epsilon$ can be made much smaller than $0.01$ in the second phase, we can achieve a large enough e-foldings.
Recent studies on the hybrid inflation after BICEP2,  one can refer to Refs.~\cite{Carrillo-Gonzalez:2014tia,Buchmuller:2014epa,Kobayashi:2014rla}.

This paper is organized as follows. In Section~\ref{double}, we briefly explain our setup and in Section~\ref{sugra}, we set up a SUSY model and show the the spectrum and its index for both scalar and tensor perturbations. We conclude in Section~\ref{sec:conclusion}.


\section{Two field inflation}
\label{double}

In this section, we briefly review a general two field inflation model with a potential separable by sum \cite{Choi:2007su}, 
\dis{
W(\phi,\chi)=U(\phi)+V(\chi). \label{sum}
} 
During the slow-roll inflation, the fields must satisfy the equations of motion,
\dis{
3H\dot{\phi} + \frac{\partial W }{\partial \phi} = 0,\qquad 3H\dot{\chi} + \frac{\partial W }{\partial \chi} = 0, 
\label{EOM}
}
respectively, and hence the fields satisfy  
\dis{
\int \frac{d \phi }{\partial W/ \partial \phi } = \int \frac{d \chi }{\partial W/ \partial \chi }. 
}
along the trajectory. 
The number of e-foldings during the inflation is given by
\dis{
N = \int H dt ,  
}
which can be expressed in terms of the fields using the field equations in \eq{EOM}.

For the separable potential in \eq{sum} of two fields, the slow-roll parameters are given by
\dis{
&\epsilon_\phi 
    = \frac{\Mp^2}{2}\left( \frac{U_\phi}{W} \right)^2 \,,~~~
\epsilon_\chi
   = \frac{\Mp^2}{2}\left( \frac{V_\chi}{W} \right)^2 \,,
\\
& \eta_{\phi}= \Mp^2 \frac{U_{\phi\phi}}{W} , \,\,~~~~~~~
 \eta_{\chi}= \Mp^2 \frac{V_{\chi\chi}}{W}\,, 
}
where the subscripts in $U$ and $V$ stand for the partial derivatives with respect to the corresponding fields. 
Using these, the cosmological observables, the power spectrum (${\cal P}_\zeta$), 
scalar spectral index ($n_\zeta$), tensor-to-scalar ratio ($r$), and its spectral index ($n_r$) can be expressed in terms of the slow-roll parameters as follows \cite{  GarciaBellido:1995qq,VW,Choi:2007su}:  
\bea
&&{\cal P}_\zeta = \frac{W_*}{24\pi^2\Mp^4}\left(\frac{u^2}{\epsilon_\phi^*}+\frac{v^2}{\epsilon_\chi^*}\right)
= \frac{W_* u^2}{24\pi^2\Mp^4\epsilon_\phi^*}\left(1+\hat{r}\right) ,
\label{power} \\
&&n_\zeta -1 = -2(\epsilon_\phi^*+\epsilon_\chi^*)
+2\frac{-2\epsilon_\phi^*
+u^2(\eta_\phi^*+\eta_\chi^*\hat{r})}{u^2(1+\hat{r})} ,
\label{spectral} \\
&&r=\frac{16}{(\frac{u^2}{\epsilon_\phi^*}+\frac{v^2}{\epsilon_\chi^*})}
=\frac{16\epsilon_\phi^*}{u^2(1+\hat{r})},\label{t/s} \\
&&n_r = -2\frac{-2\epsilon_\phi^*
+u^2(\eta_\phi^*+\eta_\chi^*\hat{r})}{u^2(1+\hat{r})} .
\label{nr}
\eea 
In the above equations, $u$, $v$, and $\hat{r}$ are defined as  
\dis{
u\equiv\frac{U_*+\widetilde{Z}^c}{W_*} ~,\qquad v\equiv\frac{V_*-\widetilde{Z}^c}{W_*} ~, \qquad \hat{r}\equiv\frac{v^2}{u^2}\frac{\epsilon_\phi^*}{\epsilon_\chi^*} ~,
}
where  
\dis{
\widetilde{Z}^c\equiv \frac{V_c\epsilon_\phi^c-U_c\epsilon_\chi^cR^{-1}}{\epsilon_\phi^c+\epsilon_\chi^cR^{-1}} ~, \qquad
R^{-1}\equiv \frac{\partial_{\phi_c} U_c}{\partial_{\phi_c} F_c}~\frac{\partial_{\chi_c} G_c}{\partial_{\chi_c} V_c} .
}
The super- or subscripts, ``$*$'' and ``$c$'' denote the values evaluated at a few Hubble times after horizon exit and the end of (the first phase of) inflation, respectively.
Here $u$ and $v$ parametrize the end effect of inflation, 
satisfying $u+v=1$ \cite{Choi:2007su}. 
$R$ shows the deviation between a hypersurface of end of  inflation, 
$F_c(\phi_c)+G_c(\chi_c)={\rm constant}$ and an equi-potential hyper surface, $U(\phi_c)+V(\chi_c)={\rm constant}$.  
$R$ is generically of order unity.  However, it can be very large or small (even negative) depending on how the inflation ends~\cite{Lyth:2005qk,Sasaki:2008uc,Naruko:2008sq,Byrnes:2008zy,Huang:2009xa,Alabidi:2010ba,Choi:2012hea}.  
From the constraint $u+v=1$, we find easily a maximum of $r$, $r\leq 16(\epsilon_\phi^* + \epsilon_\chi^*)\equiv 16\epsilon^*$~\cite{ss1}.

For $\epsilon_\phi^c\gg\epsilon_\chi^c$, $\widetilde{Z}^c$ is approximated to 
$\widetilde{Z}^c\approx V_c-W_cR^{-1}(\epsilon_\chi^c/\epsilon_\phi^c)$. 
If $U$ and $V$ are almost constant during inflation, then, $v$ and $\hat{r}$ are approximately given by $v\approx R^{-1}(\epsilon_\chi^c/\epsilon_\phi^c)$ and $\hat{r}\approx R^{-2}(\epsilon_\chi^c/\epsilon_\phi^c)^2(\epsilon_\phi^*/\epsilon_\chi^*)$, respectively.

\section{ The double hybrid inflation}
\label{sugra}
      
Let us consider the following form of the superpotential,
\dis{ \label{superPot}
W=\kappa_1S_1\left(M_1^2-\psi_1\overline{\psi}_1\right)
+\kappa_2S_2\left(M_2^2-\psi_2\overline{\psi}_2\right)+mS_1S_2,
}
The superpotential $W$ contains the inflaton fields $S_{1,2}$ and the waterfall fields, $\{\psi_{1,2},\overline{\psi}_{1,2}\}$.
While $\{S_1,S_2\}$ carry the U(1)$_R$ charges of $2$, the other superfields are neutral. 
The last term in \eq{superPot} breaks the U(1)$_R$ symmetry softly, assuming $m\ll M_{1,2}$.  
We suppose that it is the dominant U(1)$_R$ breaking term. 
In fact, $S_1\psi_2\overline{\psi}_2$ and $S_2\psi_1\overline{\psi}_1$ are also allowed in $W$. 
For simplicity of discussion, however, let us assume that their couplings are small enough.  
Then the derived potential is
\dis{ \label{V}
V&=\left|\kappa_1(M_1^2-\psi_1\overline{\psi}_1)+ mS_2 \right|^2
+\left|\kappa_2(M_2^2-\psi_2\overline{\psi}_2)+ mS_1 \right|^2 
\\
&~~+ \kappa_1^2|S_1|^2\left(|\psi_1|^2+|\overline{\psi}_1|^2\right)
+ \kappa_2^2|S_2|^2\left(|\psi_2|^2+|\overline{\psi}_2|^2\right).
}
For $|S_1|^2\gtrsim M_1^2$ and $|S_2|^2\gtrsim M_2^2$,  
the waterfall fields become stuck to the origin, $\psi_{1,2}=\overline{\psi}_{1,2}=0$, and the potential becomes dominated by a constant energy: 
\dis{ 
V_{I}&=\kappa_1^2M_1^4+\kappa_2^2M_2^4+\sqrt{2}\kappa_1M_1^2m\phi_2 +\frac{m^2}{2}\phi_2^2
+\sqrt{2}\kappa_2M_2^2m\phi_1 +\frac{m^2}{2}\phi_1^2,
\\
&\equiv \mu^4 +A_1^3\phi_2 +\frac{m^2}{2}\phi_2^2 +A_2^3\phi_1 +\frac{m^2}{2}\phi_1^2,
\label{V1}
}
where $\phi_{1,2}$ denote the real components of $S_{1,2}$ ($\equiv {\rm Re}(S_{1,2}/\sqrt{2}))$,  
and we defined $\mu^4\equiv \kappa_1^2M_1^4+\kappa_2^2M_2^4$ and $A_{1,2}^3\equiv \sqrt{2}\kappa_{1,2}M_{1,2}^2m$ for simple notations.  
Since SUSY is broken by the positive vacuum energy, the non-zero logarithmic potential can be generated \cite{FtermInf2,FtermInf}. 
We will ignore it for the first phase of inflation because of its relative smallness. 

During the first period of inflation, the two fields drive inflation with the following slow-roll parameters: 
\dis{ \label{SR}
&\epsilon_{\phi_1}
=\frac{M_P^2A_2^6}{2\mu^8}\left(1+\frac{m^2\phi_1}{A_2^3}\right)^2 ,
\qquad
\epsilon_{\phi_2}=\frac{M_P^2}{2}\left(\frac{A_1^3+m^2\phi_2}{\mu^4}\right)^2,
\\
&\eta_{\phi_1}=\eta_{\phi_2} \equiv \eta=\frac{M_P^2m^2}{\mu^4}.
}
We assume that $M_2^2\gg M_1^2$ and so  $A_2^3\gg A_1^3$. 
If $A_2^3 \gg m^2\phi_{1,2}$, then, the almost constant $\epsilon_{\phi_1}$ is dominant over $\epsilon_{\phi_2}$ for this period. In this case, the total $\epsilon$ is approximated by
\dis{
\epsilon\equiv \epsilon_{\phi_1}+ \epsilon_{\phi_2} \approx \epsilon_{\phi_1} \approx  \frac{M_P^2A_2^6}{2\mu^8}.
} 
As will be explained later, the large $A_2^3$ is necessary for the large tensor-to-scalar ratio 
and the almost constant power spectrum during the first 7 e-folds.

The first phase of inflation continues until the field $\phi_2$ arrives at $\phi_{2}^{c} \approx \sqrt{2} M_2$.
The e-folding number for this phase ($\equiv N_I$) is given in terms of the $\phi_2$ field as
\dis{ \label{N_I1}
N_I = \frac{1}{M_P^2}\int^{\phi_{2}^{*}}_{\phi_{2}^{c}} d\phi_2\frac{ \mu^4}{A_1^3+m^2\phi_2}
=\frac{1}{\eta} ~{\rm log}\left(\frac{A_1^3+m^2\phi_2^*}{A_1^3+\sqrt{2}m^2M_2}\right) .
}
During the first phase, $\phi_1$ evolves as 
\dis{ \label{N_I2}
N_I= \frac{1}{M_P^2}\int^{\phi_{1}^{*}}_{\phi_{1}^{c}} d\phi_1\frac{ \mu^4}{A_2^3+m^2\phi_1}
= \frac{1}{\eta} ~{\rm log}\left(\frac{A_2^3+m^2\phi_1^*}{A_2^3+m^2\phi_1^c}\right) 
\approx \frac{1}{\sqrt{2\epsilon_{\phi_1}}}\left(\frac{\phi_1^*-\phi_1^c}{M_P}\right) ~,
}
where $\phi_1^c$ denotes the field value of $\phi_1$ at the end of the first phase. 
Here we assumed that $A_2^3 \gg m^2\phi_1$. 
In Eqs.~(\ref{N_I1}) and (\ref{N_I2}), $\eta$ and $\epsilon$ were defined in \eq{SR}. 
As seen in \eq{N_I2}, $N_I$ cannot be large enough, if $\phi_1^*$ should be sub-Planckian. 
It is because of the large constant $A_2^3$, suppressing the logarithmic part in \eq{N_I2}.    
Hence, the $A_2^3$ needs to be turned-off in the second phase of inflation for a large enough e-folds.  


\begin{figure}[t]
  \begin{center}
  \begin{tabular}{c}
   \includegraphics[width=0.6\textwidth]{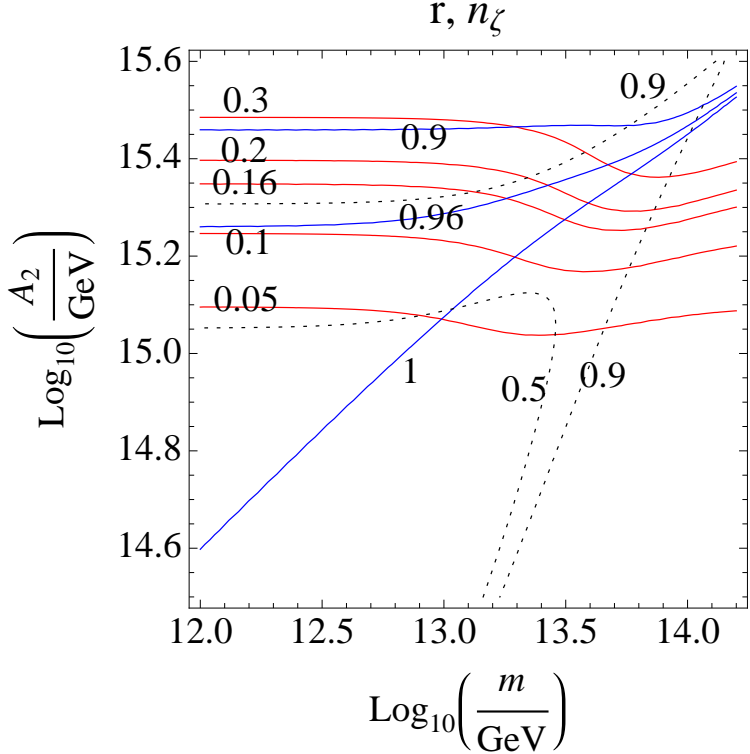}
   \end{tabular}
  \end{center}
  \caption{Contour plot of tensor-to-scalar ratio $r$ and the spectral index $n_\zeta$  in the plane of model parameters $m$ and $A_2$ at the cosmologically relevant scale. $\phi_1^*$ is adjusted to be around $N_I=7$ in \eq{N_I2}. The red lines are the contour of $r=0.05,0.1,0.16,0.2,0.3$ from the below, the blue lines are for $n_\zeta=1,0.96,0.9$ from the below and the dashed lines denote $\phi_1^*=0.5\Mp$ and $0.9\Mp$ respectively as denoted in the figure.}
\label{fig:cont1}
\end{figure}

During this period, the power spectrum is  determined by the two fields $\phi_1$  and $\phi_2$ as in \eq{power}. 
From the CMB observation, the power spectrum needs to be maintained as almost a constant 
for the first 7 e-folds corresponding to the scales $10 ~\rm{Mpc} \lesssim k^{-1} \lesssim 10^4 ~\rm{Mpc}$~\cite{Ade:2013zuv}.  
We will identify the first 7 e-folds as $N_I$.
Assuming $R^{-1}\ll 1$ and so $\hat{r}\ll 1$, we can take $u^2/\epsilon_1^* \gg v^2/\epsilon_2^*$ at the observational scale 
so that
\dis{
\calP_\zeta &
   \approx \frac{\mu^4}{24\pi^2\Mp^4\epsilon_{\phi_1}^*} ,  \\
   r &\approx 16\epsilon_{\phi_1}^*.
}
From  the observation of tensor-to-scalar ratio $r=0.16$, we can determine the scale of $\mu$:
\dis{
\mu \approx \sqrt{\kappa_2}M_2\approx  2.08\times 10^{16} \gev 
}
with $\epsilon_{\phi_1}^*\approx 0.01$.  
Since $\epsilon_{\phi_1}^*\gg\epsilon_{\phi_2}^*$ and $\hat{r}\ll 1$, the spectral index is given by
\dis{
n_\zeta \approx 1 -6\epsilon_{\phi_1}^* +2\eta_{\phi_1}^* .
}
Hence, $\eta^*_{\phi_1}=0.01$ is required for $n_\zeta\approx 0.96$. It determines $m\approx 1.8\times 10^{13} \gev$ from \eq{SR}, and $A_2\approx 2.2\times 10^{15} \gev$.
From $\epsilon\approx \epsilon_{\phi_1}^*\approx 0.01$ and $N_I\approx 7$ in \eq{N_I2}, 
we can obtain the minimum value of $\phi_1^*$, $\phi_1^*\approx 0.9  M_P$ for $\phi_1^*\gg\phi_1^c$.
On the other hand, \eq{N_I1} is easily satisfied with ${\rm log}[(A_1^3+m^2\phi_2^*)/(A_1^3+\sqrt{2}m^2M_2)]\approx 0.07$ or $\phi_2^*\gtrsim\sqrt{2}M_2$.    

In figure~\ref{fig:cont1}, we show the contour plot of the tensor-to-scalar ratio $r$ and the spectral index $n_\zeta$  in the plane of model parameters $m$ and $A_2$ at the cosmologically relevant scale. $\phi_1^*$ is adjusted to be around $N_I=7$ from \eq{N_I2}. The red lines are the contour of $r=(0.05,0.1,0.16,0.2,0.3)$ from the below, the blue lines are for $n_\zeta=(1,0.96,0.9)$ from the below and the dashed lines denote $\phi_1^*=0.5\Mp$ and $0.9\Mp$ respectively as denoted in the figure.

When $\phi_2$ reaches $\sqrt{2}M_2$, the first waterfall fields $\{\psi_2,\overline{\psi}_2\}$ become heavy and rapidly fall down to the near minima acquiring VEVs. 
$\phi_2$ also becomes heavy by the VEVs of $\{\psi_2,\overline{\psi}_2\}$ and so decoupled from the inflation.\footnote{After end of the first stage of inflation, 
the heavy fields might oscillate and affect the power spectrum as studied in Refs.~\cite{Gao:2012uq,Konieczka:2014zja}. 
In this model, however, the relevant scale is outside that can be observed by CMB and LSS. Thus, they do not affect out result.} 
As a result, 
$\phi_2$, $\kappa_2M_2^2$, and $A_2$ effectively disappear in \eq{V1}. 
Since $mS_1S_2$ term in the superpotential should also be dropped, 
the inflation is driven only by $\phi_1$ with $V_{\rm inf}=\kappa_1^2M_1^4$ after $N\approx 7$. 
In this case, we need to consider the logarithmic piece in the potential, $V_{\rm inf}\approx \kappa_1^2M_1^4\alpha {\rm log}\frac{\phi_1}{\Lambda}$, which has been neglected so far because of its smallness.      
In the second stage of inflation, thus, the potential becomes
\dis{
V_{II}&=\kappa_1^2M_1^4 \left(1+ \alpha {\rm log}\frac{\phi_1}{\Lambda} \right) ,
 \label{V2}
}
where $\alpha\approx \kappa_1^2/8\pi^2$. It is just the inflaton potential in the ordinary SUSY hybrid inflation \cite{FtermInf2,FtermInf}. 
During the second phase of inflation with the slow-roll parameters, 
\dis{
\epsilon_{II} = \frac{\alpha^2\Mp^2}{2\phi_1^2},\qquad \eta_{II} = -\frac{\alpha \Mp^2 }{\phi_1^2} ,
}
which are only relevant to smaller scales and not observable in CMB. 
The second stage of inflation continues from $\phi_1^c$ to $\phi_1^e \approx\sqrt{2} M_1$. 
The corresponding e-folding number is
\dis{
N_{II}= \frac{1}{\alpha \Mp^2 } \left[(\phi_1^c)^2 -  (\phi_1^e)^2\right].
}
With a small value of $\alpha$, therefore, we can have a sufficient e-folding number ($\sim 50$). 


So far we have not considered supergravity (SUGRA) corrections. 
Finally, we propose one example of the setups, which can protect above our discussions against SUGRA corrections. 
We suppose a logarithmic K${\rm \ddot{a}}$hler potential with a ``modulus'' $T$  
and an exponential type superpotential for stabilization of $T$: 
\dis{
K=-{\rm log}\bigg|T+T^*-\sum_i|z_i|^2\bigg|+K_X 
~~{\rm and}~~
W=W_0+W_T + W_X ,
}
where $W_T=m_{3/2}Te^{-T/f}$. 
$m_{3/2}$ and $f$ are mass parameters of order TeV and $M_P$, respectively.
Here we set $M_P=1$ for simplicity. 
While $z_i$ ($=S_{1,2}$) and $W_0$ ($=\kappa_1M_1^2S_1+\kappa_2M_2^2S_2+mS_1S_2$) are 
the fields and the superpotential {\it during inflation} considered before, 
$K_X$ and $W_X$ denote other contributions (which have not been discussed so far) to 
the K${\rm \ddot{a}}$hler and superpotential, respectively.   
Then, the $F$-term scalar potential in SUGRA is given by 
\begin{eqnarray} \label{V_sugra}
V_{F}&=&e^{K_X}\bigg[\sum_i\left|\frac{\partial W_0}{\partial z_i}\right|^2 
+\left|\frac{\partial W_T}{\partial T}\right|^2(T+T^*)
-\left\{\frac{\partial W_T}{\partial T}\left[W^*_T+W^*_X-m^*S_1^*S_2^*\right]+{\rm h.c.}\right\}
\nonumber \\
&& +\frac{1}{T+T^*-\sum_i|z_i|^2}\bigg\{\sum_{I,J}(\partial_{X_I}\partial_{X_J^*}K_X)^{-1}(D_{X_I}W)(D_{X_J}W)^*-2|W|^2\bigg\}\bigg] ,
\end{eqnarray}
where $(\partial_{X_I}\partial_{X_J^*}K_X)^{-1}$ means the inverse K${\rm \ddot{a}}$hler metric by $K_X$, and $D_{X_I}W$ is the covariant derivative in SUGRA ($=\partial W/\partial X_I+W\partial K/\partial X_I$). 
As discussed above, $S_2$ ($S_1$) is decoupled after the first (second) phase of inflation.  
The first term, $\sum_i\left|\partial W_0/\partial z_i\right|^2$ exactly reproduces \eq{V1} 
[or \eq{V} for $\{\psi_{1,2},\overline{\psi}_{1,2}\}\subset \{z_i\}$].  
It decouples from $T$ unlike the no-scale type SUGRA. 
It is because we take $-1$ as the coefficient of the logarithmic piece of the K${\rm \ddot{a}}$hler potential. 
Only if $e^{K_X}\approx 1$, thus, the SUGRA corrections leave intact \eq{V1} [or (\ref{V})].   

From the last term [and also $(\partial_{X_I}\partial_{X_J^*}K_X)^{-1}(D_{X_I}W)(D_{X_J}W)^*$] of \eq{V_sugra}, the inflaton fields potentially get Hubble scale masses during inflation. 
However, they could be smaller for $T+T^*\gtrsim 1$. 
Moreover, only if we have more fields and so e.g. 
$W= \kappa_XM_X^2X+\kappa_1M_1^2S_1+\kappa_2M_2^2S_2+mS_1S_2$ with $\kappa_XM_X^2\gtrsim  \kappa_{1,2}M_{1,2}^2$,    
then $|W|^2$ provides just a mass term of  $\kappa_XM_X^2X+\kappa_1M_1^2S_1+\kappa_2M_2^2S_2$: 
its orthogonal components, $S_1-(\kappa_1M_1^2/\kappa_XM_X^2)X$ and $S_2-(\kappa_2M_2^2/\kappa_XM_X^2)X$, 
which are approximately $S_1$ and $S_2$, respectively, still remain light enough. 
We will not discuss the dynamics of $X$ here. It would be closely associated with the complete forms of $K_X$ and $W_X$, but not directly related to our observations.

\section{Conclusion}
 \label{sec:conclusion}

The observation of B-mode polarization by BICEP2 provides hints on inflation models. 
The hybrid inflation with a single inflaton field might be difficult to accommodate all the observations within the sub-Planckian regime.
In this paper, we proposed a double hybrid inflation model, in which the inflaton potential dynamically changes with the evolution of the inflaton fields. 
During the first phase of inflation over 7 e-folds, 
the power spectrum remains almost invariant. 
The large tensor-to-scalar ratio and the constant power spectrum during the first inflationary phase are possible by a large linear term in the inflaton potential. 
In the second phase of 50 e-folds, the dominant potential becomes dynamically replaced by  the logarithmic term as in the ordinary SUSY hybrid inflation. 
Such a change in the inflaton potential is performed by the second inflaton field. 
In this model, the sub-Planckian field values ($\sim 0.9 ~M_P$) can still admit the correct cosmic observations with the sufficient e-folds.

\acknowledgments

\noindent 
K.-Y.C. appreciates Asia Pacific Center for Theoretical Physics for the support to the Topical Research Program. 
This research is supported by Basic Science Research Program through the 
National Research Foundation of Korea (NRF) funded by the Ministry of Education, 
Grant No. 2011-0011083 (K-Y.C.) and  No. 2013R1A1A2006904 (B.K.). 
B.K acknowledges the partial support 
by Korea Institute for Advanced Study (KIAS) grant funded by the Korea government.



\end{document}